\documentclass[twocolumn,aps,tightenlinefs,superscriptaddress,preprintnumbers]{revtex4}
\usepackage{graphicx,color,dcolumn,booktabs,bm}
\usepackage{longtable,lscape}
\usepackage{amsmath}
\usepackage{indentfirst}
\usepackage{epsfig}
\usepackage{feynmf}   
\usepackage{epstopdf}   
\usepackage{slashed}  
\usepackage{cases}
\usepackage{color}
\usepackage{multirow}
\usepackage{ulem}
\usepackage{verbatim}
\usepackage[colorlinks,linkcolor=red,anchorcolor=green,citecolor=blue]{hyperref}
\usepackage{mathrsfs}
\usepackage{rotating}
\usepackage{threeparttable}
\usepackage{lineno}
\usepackage{subfigure}
\usepackage{soul}

\graphicspath{{ps}}

\newcommand{\ra}{\rightarrow}
\newcommand{\EE}{e^+e^-}
\newcommand{\lamc}{\Lambda_c^+}

\newcommand{\pkpi}{pK^-\pi^+}

\newcommand{\peta}{p\eta}
\newcommand{\pomega}{p\omega}
\newcommand{\GG}{\gamma\gamma}
\newcommand{\BF}{\mathcal{B}}

\begin{document}



\title{ \quad\\[0.1cm] \boldmath Measurement of the branching fraction of $\Lambda_c^+ \to p \omega$ decay at Belle}

\noaffiliation
\affiliation{Department of Physics, University of the Basque Country UPV/EHU, 48080 Bilbao}
\affiliation{University of Bonn, 53115 Bonn}
\affiliation{Brookhaven National Laboratory, Upton, New York 11973}
\affiliation{Budker Institute of Nuclear Physics SB RAS, Novosibirsk 630090}
\affiliation{Faculty of Mathematics and Physics, Charles University, 121 16 Prague}
\affiliation{Chonnam National University, Gwangju 61186}
\affiliation{University of Cincinnati, Cincinnati, Ohio 45221}
\affiliation{Deutsches Elektronen--Synchrotron, 22607 Hamburg}
\affiliation{University of Florida, Gainesville, Florida 32611}
\affiliation{Department of Physics, Fu Jen Catholic University, Taipei 24205}
\affiliation{Key Laboratory of Nuclear Physics and Ion-beam Application (MOE) and Institute of Modern Physics, Fudan University, Shanghai 200443}
\affiliation{Justus-Liebig-Universit\"at Gie\ss{}en, 35392 Gie\ss{}en}
\affiliation{Gifu University, Gifu 501-1193}
\affiliation{SOKENDAI (The Graduate University for Advanced Studies), Hayama 240-0193}
\affiliation{Gyeongsang National University, Jinju 52828}
\affiliation{Department of Physics and Institute of Natural Sciences, Hanyang University, Seoul 04763}
\affiliation{University of Hawaii, Honolulu, Hawaii 96822}
\affiliation{High Energy Accelerator Research Organization (KEK), Tsukuba 305-0801}
\affiliation{J-PARC Branch, KEK Theory Center, High Energy Accelerator Research Organization (KEK), Tsukuba 305-0801}
\affiliation{National Research University Higher School of Economics, Moscow 101000}
\affiliation{Forschungszentrum J\"{u}lich, 52425 J\"{u}lich}
\affiliation{IKERBASQUE, Basque Foundation for Science, 48013 Bilbao}
\affiliation{Indian Institute of Science Education and Research Mohali, SAS Nagar, 140306}
\affiliation{Indian Institute of Technology Guwahati, Assam 781039}
\affiliation{Indian Institute of Technology Hyderabad, Telangana 502285}
\affiliation{Indian Institute of Technology Madras, Chennai 600036}
\affiliation{Indiana University, Bloomington, Indiana 47408}
\affiliation{Institute of High Energy Physics, Chinese Academy of Sciences, Beijing 100049}
\affiliation{Institute of High Energy Physics, Vienna 1050}
\affiliation{Institute for High Energy Physics, Protvino 142281}
\affiliation{INFN - Sezione di Napoli, I-80126 Napoli}
\affiliation{INFN - Sezione di Roma Tre, I-00146 Roma}
\affiliation{INFN - Sezione di Torino, I-10125 Torino}
\affiliation{Advanced Science Research Center, Japan Atomic Energy Agency, Naka 319-1195}
\affiliation{J. Stefan Institute, 1000 Ljubljana}
\affiliation{Institut f\"ur Experimentelle Teilchenphysik, Karlsruher Institut f\"ur Technologie, 76131 Karlsruhe}
\affiliation{Kavli Institute for the Physics and Mathematics of the Universe (WPI), University of Tokyo, Kashiwa 277-8583}
\affiliation{Department of Physics, Faculty of Science, King Abdulaziz University, Jeddah 21589}
\affiliation{Kitasato University, Sagamihara 252-0373}
\affiliation{Korea Institute of Science and Technology Information, Daejeon 34141}
\affiliation{Korea University, Seoul 02841}
\affiliation{Kyoto Sangyo University, Kyoto 603-8555}
\affiliation{Kyungpook National University, Daegu 41566}
\affiliation{P.N. Lebedev Physical Institute of the Russian Academy of Sciences, Moscow 119991}
\affiliation{Faculty of Mathematics and Physics, University of Ljubljana, 1000 Ljubljana}
\affiliation{Ludwig Maximilians University, 80539 Munich}
\affiliation{Luther College, Decorah, Iowa 52101}
\affiliation{Malaviya National Institute of Technology Jaipur, Jaipur 302017}
\affiliation{Faculty of Chemistry and Chemical Engineering, University of Maribor, 2000 Maribor}
\affiliation{Max-Planck-Institut f\"ur Physik, 80805 M\"unchen}
\affiliation{School of Physics, University of Melbourne, Victoria 3010}
\affiliation{University of Mississippi, University, Mississippi 38677}
\affiliation{University of Miyazaki, Miyazaki 889-2192}
\affiliation{Moscow Physical Engineering Institute, Moscow 115409}
\affiliation{Graduate School of Science, Nagoya University, Nagoya 464-8602}
\affiliation{Kobayashi-Maskawa Institute, Nagoya University, Nagoya 464-8602}
\affiliation{Universit\`{a} di Napoli Federico II, I-80126 Napoli}
\affiliation{Nara Women's University, Nara 630-8506}
\affiliation{National Central University, Chung-li 32054}
\affiliation{National United University, Miao Li 36003}
\affiliation{Department of Physics, National Taiwan University, Taipei 10617}
\affiliation{H. Niewodniczanski Institute of Nuclear Physics, Krakow 31-342}
\affiliation{Nippon Dental University, Niigata 951-8580}
\affiliation{Niigata University, Niigata 950-2181}
\affiliation{Novosibirsk State University, Novosibirsk 630090}
\affiliation{Osaka City University, Osaka 558-8585}
\affiliation{Pacific Northwest National Laboratory, Richland, Washington 99352}
\affiliation{Panjab University, Chandigarh 160014}
\affiliation{Peking University, Beijing 100871}
\affiliation{University of Pittsburgh, Pittsburgh, Pennsylvania 15260}
\affiliation{Punjab Agricultural University, Ludhiana 141004}
\affiliation{Research Center for Nuclear Physics, Osaka University, Osaka 567-0047}
\affiliation{Meson Science Laboratory, Cluster for Pioneering Research, RIKEN, Saitama 351-0198}
\affiliation{Dipartimento di Matematica e Fisica, Universit\`{a} di Roma Tre, I-00146 Roma}
\affiliation{Department of Modern Physics and State Key Laboratory of Particle Detection and Electronics, University of Science and Technology of China, Hefei 230026}
\affiliation{Showa Pharmaceutical University, Tokyo 194-8543}
\affiliation{Soongsil University, Seoul 06978}
\affiliation{Sungkyunkwan University, Suwon 16419}
\affiliation{School of Physics, University of Sydney, New South Wales 2006}
\affiliation{Department of Physics, Faculty of Science, University of Tabuk, Tabuk 71451}
\affiliation{Tata Institute of Fundamental Research, Mumbai 400005}
\affiliation{Department of Physics, Technische Universit\"at M\"unchen, 85748 Garching}
\affiliation{School of Physics and Astronomy, Tel Aviv University, Tel Aviv 69978}
\affiliation{Toho University, Funabashi 274-8510}
\affiliation{Department of Physics, Tohoku University, Sendai 980-8578}
\affiliation{Earthquake Research Institute, University of Tokyo, Tokyo 113-0032}
\affiliation{Department of Physics, University of Tokyo, Tokyo 113-0033}
\affiliation{Tokyo Institute of Technology, Tokyo 152-8550}
\affiliation{Tokyo Metropolitan University, Tokyo 192-0397}
\affiliation{Virginia Polytechnic Institute and State University, Blacksburg, Virginia 24061}
\affiliation{Wayne State University, Detroit, Michigan 48202}
\affiliation{Yamagata University, Yamagata 990-8560}
\affiliation{Yonsei University, Seoul 03722}
  \author{S.~X.~Li}\affiliation{Key Laboratory of Nuclear Physics and Ion-beam Application (MOE) and Institute of Modern Physics, Fudan University, Shanghai 200443} 
  \author{L.~K.~Li}\affiliation{University of Cincinnati, Cincinnati, Ohio 45221} 
  \author{C.~P.~Shen}\affiliation{Key Laboratory of Nuclear Physics and Ion-beam Application (MOE) and Institute of Modern Physics, Fudan University, Shanghai 200443} 
  \author{I.~Adachi}\affiliation{High Energy Accelerator Research Organization (KEK), Tsukuba 305-0801}\affiliation{SOKENDAI (The Graduate University for Advanced Studies), Hayama 240-0193} 
  \author{H.~Aihara}\affiliation{Department of Physics, University of Tokyo, Tokyo 113-0033} 
  \author{S.~Al~Said}\affiliation{Department of Physics, Faculty of Science, University of Tabuk, Tabuk 71451}\affiliation{Department of Physics, Faculty of Science, King Abdulaziz University, Jeddah 21589} 
  \author{D.~M.~Asner}\affiliation{Brookhaven National Laboratory, Upton, New York 11973} 
  \author{T.~Aushev}\affiliation{National Research University Higher School of Economics, Moscow 101000} 
  \author{P.~Behera}\affiliation{Indian Institute of Technology Madras, Chennai 600036} 
  \author{K.~Belous}\affiliation{Institute for High Energy Physics, Protvino 142281} 
% ...
\collaboration{The Belle Collaboration}

\begin{abstract}
     Using 980.6 $\rm fb^{-1}$ of data collected with the Belle detector operating at the KEKB asymmetric-energy $e^+e^-$ collider, we present a measurement of the branching fraction of the singly Cabibbo-suppressed decay $\Lambda_c^+ \to p \omega$. A clear $\Lambda_c^+$ signal is observed for $\Lambda_c^+ \to p \omega$ with a statistical significance of 9.1 standard deviations, and we measure the ratio of branching fractions ${\cal B}(\Lambda_c^+ \to p \omega)/{\cal B}(\Lambda_c^+ \to p K^- \pi^+) =  (1.32 \pm 0.12 (\rm stat) \pm 0.10 (\rm syst))\times 10^{-2}$, from which we infer the branching fraction ${\cal B}(\Lambda_c^+ \to p \omega) = (8.27 \pm 0.75 (\rm stat) \pm 0.62 (\rm syst) \pm 0.42 (\rm ref))\times 10^{-4}$. The first quoted uncertainty is statistical, the second systematic, and the third from the reference mode $\Lambda_c^+ \to p K^- \pi^+$.
\end{abstract}

\maketitle

\tighten

\section{\boldmath Introduction}
\setstcolor{red}
     Charmed mesons and baryons are copiously produced in the B-factory experiment, providing an excellent arena for understanding Quantum Chromodynamics (QCD) with transitions involving charm quark.
     SU(3)$_F$ flavor symmetry~\cite{theoC1, theoC2} and QCD dynamical models~\cite{theoD1, theoD2, theoD3} provide theoretical estimates of charmed baryon decays. The former relies on experimental results as the input; the latter models often make different predictions for unknown baryon wave functions and non-factorizable contributions, which makes it difficult to perform definitive tests between theoretical models.

     Experimentally, the investigation of charmed baryon decays is more difficult than for charmed mesons due to their smaller production rate. Only the lowest-lying charmed baryon $\lamc$ decays weakly.
     Since it was first discovered~\cite{theoA1}, many hadronic weak decays, mostly Cabibbo-favored, have been observed~\cite{pdg}. In contrast, the knowledge of Cabibbo-suppressed decays has been limited. Both measurements and theoretical models point to non-factorizable contributions, such as $W$-exchange, having a sizable impact on individual decay rates as well as the total widths~\cite{theoB1, theoB2, theoB3, theoB4}.

     Recently, the LHCb Collaboration reported the first observation of a singly Cabibbo-suppressed (SCS) decay $\lamc \ra \pomega (\ra \mu^+\mu^-)$ with a statistical significance of five standard deviations ($\sigma$). They measured a branching fraction value of $\BF(\lamc \ra \pomega) = 9.4\pm3.9)\times 10^{-4}$~\cite{lhcbpomega}.
     Theoretical predictions exist, for this particular decay, based either on SU(3)$_F$ flavor symmetry~\cite{theoE1,theoE2} or QCD dynamical model predictions~\cite{theoF1}.

     In this analysis, we measure the branching fraction of the $\lamc \ra \pomega (\ra \pi^+\pi^-\pi^0)$ channel for the first time at Belle, taking advantage of the large value of $\BF(\omega \ra \pi^+\pi^-\pi^0)$~\cite{pdg}.
     To improve the measurement precision, we measure the ratio of the branching fractions of this SCS process with respect to the $\lamc \ra \pkpi$ reference decay mode:
        \begin{equation}
        \label{bf-cal}
           \frac{\BF(\lamc \ra \pomega)}{\BF(\lamc \ra \pkpi)} = \frac{N^{\rm data}_{\rm sig} \times \epsilon^{\rm MC}_{\rm ref}}{N^{\rm data}_{\rm ref}\times \epsilon^{\rm MC}_{\rm sig}\times \BF'},
        \end{equation}
     where $N^{\rm data}$ and $\epsilon^{\rm MC}$ are the number of fitted $\Lambda_c^+$ events in data and the detection efficiency, respectively; the subscript ``ref'' refers to the reference mode and ``sig'' to the signal mode; and $\BF'=\BF(\omega \ra \pi^+ \pi^- \pi^0) \times \BF(\pi^0 \ra \GG)$~\cite{pdg}.

\section{\boldmath The data sample and the belle detector}

     Measurement of the branching fraction of $\lamc \ra \pomega$ is based on a data sample taken at or near the $\Upsilon(1S)$, $\Upsilon(2S)$, $\Upsilon(3S)$, $\Upsilon(4S)$, and $\Upsilon(5S)$ resonances collected with the Belle detector at the KEKB asymmetric-energy $e^+e^-$ collider~\cite{KEKB},  corresponding to an integrated luminosity of 980.6 $\rm fb^{-1}$.
     The Belle detector is a large-solid-angle magnetic spectrometer that consists of a silicon vertex detector (SVD), a 50-layer central drift chamber (CDC), an array of aerogel threshold Cherenkov counters (ACC), a barrel-like arrangement of time-of-flight scintillation counters (TOF), and an electromagnetic calorimeter comprised of CsI(Tl) crystals (ECL) located inside a superconducting solenoid coil that provides a 1.5~T magnetic field. An iron flux-return located outside of the coil is instrumented to detect $K_L^0$ mesons and to identify muons (KLM).  The detector is described in detail elsewhere~\cite{Belle}.

     A signal Monte Carlo (MC) sample of $\EE \ra c\bar{c}$; $c\bar{c} \ra \lamc X$ with $X$ denoting anything; $\lamc \ra \pomega$ with $\omega \ra \pi^+\pi^-\pi^0$, $\pi^0 \ra \GG$ is used to optimize the selection criteria and estimate the reconstruction and selection efficiency. Events are generated with {\sc pythia}~\cite{pythia} and {\sc EvtGen}~\cite{evtgen}, and decay products are propagated by {\sc geant3}~\cite{geant3} to simulate the detector performance.  Charge-conjugate modes are also implied unless otherwise stated throughout this paper.

     Inclusive MC samples of $\Upsilon(4S)\ra B^{+}B^{-}/B^{0}\bar{B}^{0}$, $\Upsilon(5S)\ra B_{s}^{(*)}\bar{B}_{s}^{(*)}$, $\EE \to q\bar{q}$ $(q=u,~d,~s,~c)$ at $\sqrt{s}$ = 10.52, 10.58 and 10.867~GeV, and $\Upsilon(1S,~2S,~3S)$ decays, corresponding to four times the integrated luminosity of each data set, are used to characterize the backgrounds~\cite{topoana}.

\section{\boldmath Event selection}
     The $\lamc$ candidates are reconstructed in two decay modes, $\lamc \ra \pkpi$ and $\lamc \ra \pomega$ with $\omega \ra \pi^+\pi^-\pi^0$, $\pi^0 \ra \GG$, corresponding to the reference and signal modes, respectively. Final-state charged particles, $p$, $K$, and $\pi$, are selected using the likelihood information derived from the charged-hadron identification systems (ACC, TOF, CDC) into a combined likelihood, $\mathcal{R}(h|h') = \mathcal{L}(h)/(\mathcal{L}(h)+\mathcal{L}(h'))$ where $h$ and $h'$ are $\pi$, $K$, and $p$ as appropriate~\cite{pidcode}. The protons are required to have $\mathcal{R}(p|\pi)>0.9$ and $\mathcal{R}(p|K)>0.9$, charged kaons to have $\mathcal{R}(K|p) > 0.4$ and $\mathcal{R}(K|\pi) > 0.9$, and charged pions to have $\mathcal{R}(\pi|p) > 0.4$ and $\mathcal{R}(\pi|K) > 0.4$.  A likelihood ratio for $e$ and $h$ identification, $\mathcal{R}(e)$, is formed from ACC, CDC, and ECL information~\cite{eidcode}, and is required to be less than 0.9 for all charged tracks to suppress electrons. For the typical momentum range of our signal decay, the identification efficiencies of $p$, $K$, and $\pi$ are 82\%, 70\%, and 97\%, respectively.
     Probabilities of misidentifying $h$ as $h'$, $P(h\ra h')$, are estimated 
     to be 3\% [$P(p\ra \pi)$], 7\% [$P(p\ra K)$], 10\% [$P(K\ra \pi)$], 2\% [$P(K\ra p)$], 5\% [$P(\pi\ra K)$], and 1\% [$P(\pi\ra p)$].
     Furthermore, for each charged-particle track, the distance of closest approach with respect to the interaction point along the direction opposite the $e^+$ beam ($z$ axis) and in the transverse $r\phi$ plane is required to be less than 2.0 cm and 0.1 cm, respectively. In addition, at least one SVD hit for each track is required.

     For $\lamc \ra \pkpi$, a common vertex fit is performed on $\lamc$ candidates and the corresponding $\chi^2_{\rm vtx}$ value is required to be less than 40 to reject the combinatorial background.
     We require a scaled momentum of $x_p > 0.53$ to suppress the background, especially from $B$-meson decays, where $x_{p} = {p^{*}}/{\sqrt{E^{2}_{\rm cm}/4 - M^{2}}}$~\cite{speedoflight}, and $E_{\rm cm}$ is the center-of-mass (CM) energy, $p^{*}$ and $M$ are the momentum and invariant mass, respectively, of the $\Lambda_c^+$ candidates in the CM frame. All of these optimized selection criteria are the same as those in our previous publication~\cite{ana-peta-ppi0}.

     An ECL cluster not matching any track is identified as a photon candidate.
     To reject neutral hadrons, the sum of the energy deposited in the central $3\times3$ square of ECL cells is required to be at least 90\% of the total energy deposited in the enclosing $5\times5$ square of cells for each photon candidate. Moreover, the energy of photon candidates must exceed 50~MeV and 70~MeV in the barrel ($-0.63<$ $\cos\theta$ $<0.85$) and endcap ($-0.91<$ $\cos \theta < -0.63$ or $0.85<$ $\cos \theta<0.98$) regions of the ECL, respectively, where $\theta$ is the inclination angle with respect to the $z$ axis. A $\pi^0$ candidate is reconstructed by two photons and $0.08<M(\GG)<0.18$ GeV/$c^2$ is required. We perform a mass-constrained (1C) fit on the two photons to require their mass at the $\pi^0$ nominal mass~\cite{pdg} and the corresponding $\chi^2_{\rm 1C}$ value must be less than 10. For $\omega \ra \pi^+\pi^-\pi^0$, we place a requirement on the momentum of $\omega$ candidates in the CM frame: $P^{*}(\omega) > 0.9$ GeV/c. An $\omega$ candidate and a proton candidate are combined to form a $\Lambda_c^+$ candidate. A common vertex fit is performed for the three charged tracks, $p$ and $\pi^{\pm}$, and the requirement of $\chi^2_{\rm vtx}<15$ is set to suppress background events without a common vertex, especially due to long-lived particles such as $K_S^0$ and $\Sigma^+$. Again, $x_p>0.53$ is required for $\lamc \ra \pomega$ candidates.
     With the above requirements, $\sim$8\% of events have multiple $\lamc$ candidates. We select the best $\lamc$ candidate based on the minimum $\chi^2_{1C}$ value;  the efficiency for this best candidate selection is around 70\%. All the above selection criteria are based on an optimization with a maximum figure-of-merit $S/\sqrt{S+B}$, where $S$ and $B$ are the numbers of signal and background events, respectively, expected in the $\lamc$ signal region [(2.25, 2.32) GeV/$c^2$, corresponding to $\pm2.5\sigma$ around the nominal $\lamc$ mass~\cite{pdg}].
     $S$ is estimated via $\frac{N^{\rm data}_{\rm ref}\times \epsilon^{\rm MC}_{\rm sig}\times \BF'}{\epsilon^{\rm MC}_{\rm ref}}\times \frac{\BF(\lamc \ra \pomega)}{\BF(\lamc \ra \pkpi)}$, where $\BF(\lamc \ra \pomega)$ is assumed to be $9.4\times10^{-4}$~\cite{lhcbpomega}, while the other parameters have been introduced in Eq.~(\ref{bf-cal}). Likewise, $B$ is the number of background events obtained from inclusive MC samples normalized to the signal region.

     From the study of inclusive MC samples~\cite{topoana}, there are several peaking backgrounds from the decays $\lamc \ra K_S^0 p \pi^0$ with $K_S^0 \ra \pi^+\pi^-$, $\lamc \ra \Sigma^+ \pi^+\pi^-$ with $\Sigma^+ \ra p \pi^0$, $\lamc \ra \Lambda \pi^+ \pi^0$ with $\Lambda \ra p\pi^-$, and $\lamc \ra \Delta^{++} \pi^-\pi^0$ with $\Delta^{++} \ra p \pi^+$, which have the same final-state topology as the signal. However, owing to the long lifetime of $K_S^0$, $\Sigma^+$, and $\Lambda$, many of the decay vertices of these particles are displaced by several centimeters from the main vertex.
     Therefore, the $\chi^2_{\rm vtx}$ requirement suppresses most of these background events, subsequently leaving no $K_S^0$ nor $\Sigma^+$ peaks in the $M(\pi^+\pi^-)$ and $M(p\pi^0)$ distributions, respectively.
     In the $M(p\pi^-)$ spectrum, a $\Lambda$ signal is seen and an optimized requirement of $|M(p\pi^-)-m(\Lambda)|>2.756$ MeV/$c^2$ ($\approx$ 3$\sigma$) is placed, where $m(\Lambda)$ is the nominal mass of $\Lambda$~\cite{pdg}. There is a small $\Delta^{++}$ signal observed in the $M(p\pi^+)$ distribution. Due to the broad width of the $\Delta^{++}$ ($\sim$118 MeV)~\cite{pdg}, no requirement on $M(p\pi^+)$ is imposed. Since such a background can be described by the $\omega$ sidebands, a simultaneous fit to the $M(p\omega)$ distributions from the selected events in the $\omega$ signal region and the normalized $\omega$ sidebands is used to handle the $\Delta^{++}$ background in extracting the $\lamc$ signal events, as introduced in the following section.

\section{\boldmath efficiency estimation and fit results}
\begin{figure}[h!]
             \centering
             \includegraphics[width=0.4\textwidth]{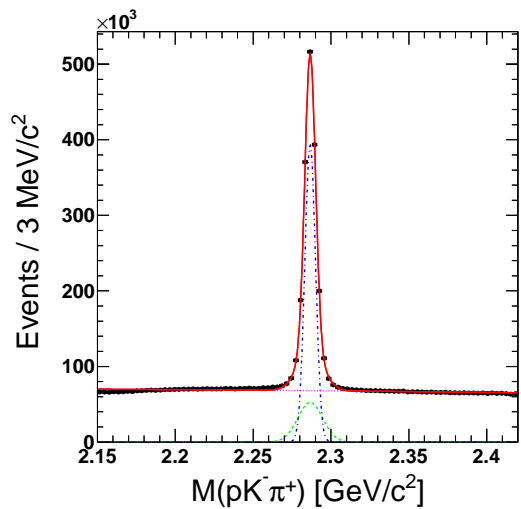}
             \caption{Fit to the invariant mass distribution of $\pkpi$ from data. Black dots with error bars represent the data; the pink dashed line, the blue dash-dotted line, the green long-dashed line, and the red solid line represent the background contribution, the core Gaussian, tail Gaussian, and the total fit, respectively.}
             \label{pkpi-data}
     \end{figure}

     To measure the ratio of the branching fractions, ${\cal B}(\lamc \ra \pomega)/{\cal B}(\lamc \ra \pkpi)$, we first determine the yields of $\lamc \ra \pkpi$ and $\lamc \ra p\omega$ by fitting the corresponding invariant mass distributions. Figure~\ref{pkpi-data} shows the $M(\pkpi)$ distribution overlaid with the fit result. A clear $\lamc$ signal is seen and we fit the $M(\pkpi)$ distribution using a binned maximum likelihood fit with a bin width of 3 MeV/$c^{2}$. A sum of two Gaussian functions with a common mean value is used to model the signal events and a second-order polynomial is used to model the background events. The parameters of the signal and background shapes are free in the fit. The reduced $\chi^{2}$ value of the fit is $\chi^2/{\rm ndf}=87/82=1.06$ and the fitted number of signal events is 1476200 $\pm$ 1560, where $\rm ndf$ is the number of degrees of freedom and the uncertainty is statistical only. The signal efficiency for this reference mode is estimated to be $(14.06\pm 0.01)$\% via a Dalitz-plot method~\cite{dalitz}; the details can be found in Ref.~\cite{ana-peta-ppi0}.

     Since the decay $\lamc \ra p\eta$ with $\eta \ra \GG$ has been well measured~\cite{ana-peta-ppi0}, the same transition $\lamc \ra p\eta$, followed by the decay $\eta \ra \pi^+\pi^-\pi^0$, having the same final-state topology as our signal mode, is taken as a control channel to validate the event selection criteria. With the final selection criteria, a clear $\eta$ signal is observed in the $M(\pi^+\pi^-\pi^0)$ distribution and the $\eta$ signal region is defined as 0.535 $\textless$ $M(\pi^+\pi^-\pi^0)$ $\textless$ 0.561 GeV/$c^2$. In the $M(\peta)$ distribution, a significant $\lamc$ signal is observed and a one-dimensional fit is performed on the $M(\peta)$ distribution using an unbinned maximum-likelihood method. A sum of two Gaussian functions with the same mean value is used to model the $\lamc$ signal and a second-order polynomial function is used to model the background, with all parameters floated in the fit. The determined number of $\lamc$ signal events is $819.9\pm78.6$ and the signal efficiency is $(1.48\pm 0.01)$\%, as determined from a signal MC sample. Therefore, the branching ratio of $\lamc \ra p\eta$ with respect to the reference mode $\lamc \ra \pkpi$ is $\frac{\BF(\lamc \ra \peta)}{\BF(\lamc \ra \pkpi)} = 0.0233 \pm 0.0022$, resulting in the branching fraction $\BF(\lamc \ra \peta) = (1.46\pm0.14) \times 10^{-3}$, where the uncertainty is statistical only. Comparing with the result of a previous dedicated measurement, $\BF(\lamc \ra \peta)=(1.42\pm0.05(\rm stat)\pm0.11(\rm syst))\times10^{-3}$~\cite{ana-peta-ppi0}, we find they are consistent with each other.

\begin{figure}[h!]
             \centering
             \includegraphics[width=0.4\textwidth]{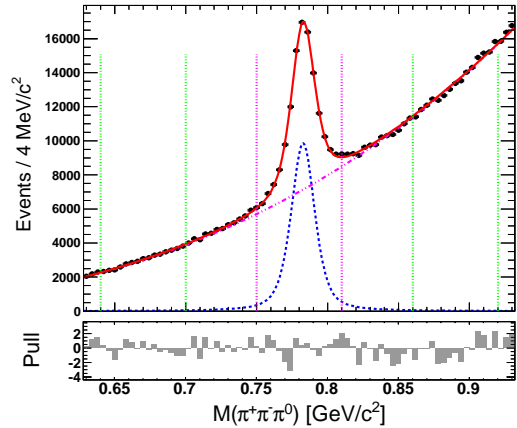}
             \caption{A fit to the $\pi^+\pi^-\pi^0$ invariant mass distribution is shown. The black dots with error bars represent the data; the red solid line represents the total fitted result; the blue dashed line represents the signal shape; and the magenta dashed-dotted line represents the fitted background. The region between the two violet vertical lines is regarded as the signal region and the two regions between the pairs of green vertical lines are regarded as the $\omega$ sideband regions.}
             \label{data-fit-omegarange}
     \end{figure}

     With the final selection criteria applied, the $\pi^+\pi^-\pi^0$ invariant mass distribution is displayed in Fig.~\ref{data-fit-omegarange}. There is a clear $\omega$ signal and a fit to the sum of a polynomial and a signal function is performed using an unbinned maximum-likelihood method. The $\omega$ signal is described by a Breit-Wigner (BW) function convolved with a double Gaussian function to represent the detector resolution. The mass and width of the BW function are set to the $\omega$ world average value~\cite{pdg}, the means are constrained to be the same for the double Gaussian function, and the remaining parameters are free.
     A third-order polynomial function is used to model the combinatorial background. The fit result is shown in Fig.~\ref{data-fit-omegarange}, along with the pulls $(N_{\rm data}-N_{\rm fit})/\sigma_{\rm data}$, where $\sigma_{\rm data}$ is the error on $N_{\rm data}$. The $\omega$ signal region is determined to be 0.75 to 0.81 GeV/$c^2$ in the $M(\pi^+\pi^-\pi^0)$ spectrum, corresponding to a 92\% selection efficiency, and the sideband regions of $\omega$ are set to be (0.64, 0.70) GeV/$c^2$ and (0.86, 0.92) GeV/$c^2$.

\begin{figure}[h!]
             \centering
             \includegraphics[width=0.4\textwidth]{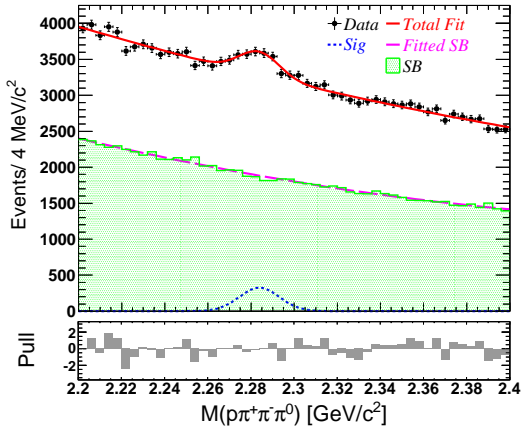}
             \caption{A simultaneous fit to the $p\omega$ invariant mass distribution in the $\omega$ signal region, and the normalized $\omega$ sideband regions is shown. The black dots with error bars represent the data; the red solid line represents the total fitted result; the blue dashed line represents the signal shape; the magenta long-dashed line represents the fitted sideband line shape, and the green filled region is from the normalized sideband regions.}
             \label{data-fit-mpomega}
     \end{figure}

     The $M(p\omega)$ distribution for events in the $\omega$ signal region and the normalized $\omega$ sideband regions are shown in Fig.~\ref{data-fit-mpomega}.
     There is a clear $\lamc$ signal observed and we perform a simultaneous extended unbinned maximum-likelihood fit to extract the $\lamc$ signal yield.
     The function for an event in the $\omega$ signal region (SR) is described as
          \begin{equation}
             \label{pdfsr-cal}
             \begin{split}
             F_{\rm sr}(M_i) &= n_{\rm s} \mathcal{P}_{\rm s}(M_i) + n_{\rm b} \mathcal{P}_{\rm b}(M_i) \\
                    & + f_{\rm norm}[n_{\rm sb}^{\rm s} \mathcal{P}_{\rm sb}^{\rm s}(M_i) + n_{\rm sb}^{\rm b} \mathcal{P}_{\rm sb}^{\rm b}(M_i)]
             \end{split}
          \end{equation}
          and that for an event in the $\omega$ sidebands (SB) is
          \begin{equation}
             \label{pdfsb-cal}
             F_{\rm sb}(M_j) = n_{\rm sb}^{\rm s} \mathcal{P}_{\rm sb}^{\rm s}(M_j) + n_{\rm sb}^{\rm b} \mathcal{P}_{\rm sb}^{\rm b}(M_j),
          \end{equation}
          where $\mathcal{P}_{\rm s}$ and $\mathcal{P}_{\rm b}$ are probability density functions (PDFs) of the $\lamc$ signal and background for the $M(p\omega)$ distribution with the events in SR, respectively; $\mathcal{P}_{\rm sb}^{\rm s}$ and $\mathcal{P}_{\rm sb}^{\rm b}$ are the $\lamc$ signal and background PDFs for the $M(p\omega)$ distribution with the events in SB; $n_{\rm s}$, $n_{\rm b}$, $n_{\rm sb}^{\rm s}$, and $n_{\rm sb}^{\rm b}$ are the corresponding numbers of the fitted events; $f_{\rm norm}=S_{\rm sb}/S_{\rm sr}=0.428$ is the normalization factor determined by fitting the $M(\pi^+\pi^-\pi^0)$ distribution ($S_{\rm sb}$ and $S_{\rm sr}$ are the numbers of the fitted background events in defined $\omega$ sidebands and signal region, respectively). The extended likelihood function is
          \begin{equation}
             \label{LL-cal}
                \mathcal{L}=\frac{e^{-n_{\rm sr}}}{N_{\rm sr}!}\prod_i^{N_{\rm sr}} F_{\rm sr}(M_i) \frac{e^{-n_{\rm sb}}}{N_{\rm sb}!} \prod_j^{N_{\rm sb}} F_{\rm sb}(M_j),
          \end{equation}
          where $n_{\rm sr} = n_{\rm s} + n_{\rm b} + f_{\rm norm}(n_{\rm sb}^{\rm s} + n_{\rm sb}^{\rm b})$, $n_{\rm sb} = n_{\rm sb}^{\rm s} + n_{\rm sb}^{\rm b}$, and $N_{\rm sr}$ and $N_{\rm sb}$ are the number of events in SR and SB.
     The $\mathcal{P}_{\rm s}$ and $\mathcal{P}_{\rm sb}^{\rm s}$ are both a sum of two Gaussian functions with the same mean value. The parameters of $\mathcal{P}_{\rm s}$ and $\mathcal{P}_{\rm sb}^{\rm s}$ are kept the same and floated. The $\mathcal{P}_{\rm b}$ and $\mathcal{P}_{\rm sb}^{\rm b}$ are described by second-order and third-order polynomial functions, respectively. All parameters of the background functions are free. The fit result and pulls are shown in Fig.~\ref{data-fit-mpomega}. After fitting, $n_{\rm s} = 1829\pm168$ and $n^{\rm s}_{\rm sb}=39\pm14$ are obtained. The $\chi^2/\rm ndf$ for the fit is $44/41 = 1.07$ for the fit. The statistical significance is evaluated with $\sqrt{-2 \rm ln(\mathcal{L}_0/\mathcal{L}_{\rm max})}$, where $\mathcal{L}_0$ is the maximized likelihood value with the number of signal events set to zero, and $\mathcal{L}_{\rm max}$ is the nominal maximized likelihood value. We obtain 9.1$\sigma$ as the statistical significance.

     With all event selections, the $M(p\omega)$ distribution from signal MC sample is obtained and signal events of $\lamc$ are determined by fitting the $M(p\omega)$ distribution. We use a sum of two Gaussian functions with the same mean value to model the signal and a second-order polynomial function to model the background. All parameters of the signal and background functions are free.
     The efficiency of our signal decay is obtained by the ratio of the number of fitted signal events in the $M(p\omega)$ distribution to that of generated events from signal MC sample, which is $(1.50 \pm 0.01)$\%, where the uncertainty is statistical only. The branching ratio is thus $\BF(\lamc \ra \pomega)/\BF(\lamc \ra \pkpi) = (1.32 \pm 0.12)\times10^{-2}$, where the uncertainty is statistical.

\section{\boldmath Systematic Uncertainties}
     Since the branching fraction is obtained from a ratio of quantities in Eq.~(\ref{bf-cal}), some systematic uncertainties cancel. The sources of systematic uncertainties include the fits of the reference and signal modes, particle identification (PID), photon efficiency, the uncertainties of branching fractions for the $\omega \ra \pi^+\pi^-\pi^0$ and $\pi^{0} \ra \GG$ decays, and the statistics of the signal MC sample.

     The systematic uncertainty from the fit of the $M(\pkpi)$ spectrum is estimated by modifying the signal and background functions, bin width, and the fit range. To evaluate the uncertainty from the signal function, the signal shape is fixed to that from the fit to the MC sample. The uncertainty from the background shape is assessed by using a  first-order polynomial. Furthermore, the bin width is varied from 2 MeV/$c^2$ to 4 MeV/$c^2$, and the fit range of the invariant mass spectrum adjusted to estimate the uncertainties from binning and fit range. The fractional difference in measured branching ratios, $2.1$\%, is taken as the uncertainty.
     The systematic uncertainty from the fit of the $M(p\omega)$ distribution is estimated by changing the signal and background line shapes, the fit range, and the fit method. The signal shape is changed from the double Gaussian function to a single Gaussian function, and the background line shape is changed from the second-order polynomial function to a third-order polynomial function, as well as enlarging the fit range. In addition, a two-dimensional unbinned maximum-likelihood fit of the ($M(p\omega)$, $M(\pi^+\pi^-\pi^0)$) distribution is performed, to evaluate the the fit method uncertainty, and the fractional difference in the branching ratio, $5.2$\%, is taken as the systematic uncertainty.

     Systematic uncertainties from PID efficiencies of the $p$ and $\pi^+$ cancel approximately, resulting in negligible amount of systematic uncertainty in the ratio. Systematic uncertainties of $1.6$\% and $1.3$\% are assigned for the $K^-$ and $\pi^-$ identification efficiencies, respectively, calculated using a $D^{*+} \ra D^{0}\pi^+ $ with $D^0 \ra K^-\pi^+$ sample. The total systematic uncertainty from PID is 2.9\%.
     The systematic uncertainty due to tracking efficiency cancels in the ratio.
     Based on a study of radiative Bhabha events, a systematic uncertainty of 2.0\% is assigned to the photon efficiency for each photon, and the total systematic uncertainty from photon reconstruction is thus 4.0\%. Since the signal efficiency is independent of the decay angular distribution of proton in the $\lamc$ rest frame, the model-dependent uncertainty has negligible effect on efficiency.
     The systematic uncertainty from $\BF(\omega \ra \pi^+\pi^-\pi^0)\times\BF(\pi^{0} \ra \GG)$ is 0.7\%~\cite{pdg}, and that from the size of the signal MC sample is estimated to be 0.8\% for $\lamc \to \pomega$.

     These systematic uncertainties are summarized in Table~\ref{error-total}, where a total systematic uncertainty of 7.6\% is obtained by assuming all uncertainties are independent and adding them in quadrature.
     \begin{table}[h]
             \centering
             \caption{Tabulation of the sources of the relative systematic uncertainties (\%) on the ratio of the branching fractions $\BF(\lamc \ra \pomega)/\BF(\lamc \ra \pkpi)$.}
             \label{error-total}
             \linespread{1.2}
             \begin{tabular}{ c  c   }   \hline \hline
                         Source  & Uncertainty (\%) \\ \hline
                         Fit of reference mode                                         &2.1    \\
                         Fit of signal mode                                            &5.2    \\
                         PID                                                           &2.9    \\
                         Photon efficiency                                             &4.0    \\
                         $\BF(\omega \ra \pi^+\pi^-\pi^0)$ and $\BF(\pi^{0} \ra \GG)$  &0.7    \\
                         Statistics of signal MC sample                               &0.8    \\ \hline
                         Total                                                         &7.6    \\ \hline \hline
             \end{tabular}
      \end{table}

\section{\boldmath Result}
     We measure the ratio of branching fractions
     \begin{equation}
          \begin{split}
          \frac{\BF(\lamc \ra \pomega)}{\BF(\lamc \ra \pkpi)} = (1.32 \pm 0.12 \pm 0.10)\times10^{-2}.
          \end{split}
     \end{equation}
     Using $\BF(\lamc \ra \pkpi) = (6.28\pm0.32)\times10^{-2}$~\cite{pdg}, we obtain the branching fraction:
     \begin{equation}
          \begin{split}
          \BF(\lamc \ra \pomega) = (8.27 \pm 0.75 \pm 0.62 \pm 0.42)\times 10^{-4},
          \end{split}
     \end{equation}
     where the first uncertainty is statistical, the second systematic, and the third from the reference mode $\lamc \ra \pkpi$. This result is consistent with the LHCb result $(9.4\pm3.9)\times 10^{-4}$~\cite{lhcbpomega}, and agrees with the theoretical predictions of $(11.4\pm5.4)\times10^{-4}$~\cite{theoE1} and $(6.3\pm3.4)\times10^{-4}$~\cite{theoE2} within uncertainties based on the SU(3)$_F$ flavor symmetry. However, our result contradicts the QCD dynamical model prediction of $(3.4-3.8)\times 10^{-4}$~\cite{theoF1}. 

\section{\boldmath Conclusion}
     To conclude, we perform a measurement of the decay $\lamc \ra \pomega$ with the full Belle dataset for the first time at Belle. A $\lamc$ signal is observed in the $M(p\omega)$ distribution with a statistical significance of 9.1 standard deviations. The measured branching ratio is $\frac{\BF(\lamc \ra \pomega)}{\BF(\lamc \ra \pkpi)} = (1.32 \pm 0.12 (\rm stat) \pm 0.10 (\rm syst))\times10^{-2}$. With the independently measured value of $\BF(\lamc \ra \pkpi)$~\cite{pdg}, we extract a branching fraction of $\BF(\lamc \ra \pomega) = (8.27 \pm 0.75 (\rm stat) \pm 0.62 (\rm syst) \pm 0.42 (\rm ref))\times 10^{-4}$, where the uncertainties are statistical, systematic, and from $\BF(\lamc \ra \pkpi)$, respectively. The measured result is consistent with the LHCb result~\cite{lhcbpomega} but with a considerably improved precision.

\section{\boldmath ACKNOWLEDGMENTS}

We thank the KEKB group for the excellent operation of the
accelerator; the KEK cryogenics group for the efficient
operation of the solenoid; and the KEK computer group, and the Pacific Northwest National
Laboratory (PNNL) Environmental Molecular Sciences Laboratory (EMSL)
computing group for strong computing support; and the National
Institute of Informatics, and Science Information NETwork 5 (SINET5) for
valuable network support.  We acknowledge support from
the Ministry of Education, Culture, Sports, Science, and
Technology (MEXT) of Japan, the Japan Society for the
Promotion of Science (JSPS), and the Tau-Lepton Physics
Research Center of Nagoya University;
the Australian Research Council including grants
DP180102629, 
DP170102389, 
DP170102204, 
DP150103061, 
FT130100303; 
Austrian Science Fund (FWF);
the National Natural Science Foundation of China under Contracts
No.~11435013,  
No.~11475187,  
No.~11521505,  
No.~11575017,  
No.~11675166,  
No.~11705209;  
No.~11761141009;
No.~11975076;
No.~12042509;
No.~12135005;
Key Research Program of Frontier Sciences, Chinese Academy of Sciences (CAS), Grant No.~QYZDJ-SSW-SLH011; 
the  CAS Center for Excellence in Particle Physics (CCEPP); 
the Shanghai Pujiang Program under Grant No.~18PJ1401000;  
the Ministry of Education, Youth and Sports of the Czech
Republic under Contract No.~LTT17020;
the Carl Zeiss Foundation, the Deutsche Forschungsgemeinschaft, the
Excellence Cluster Universe, and the VolkswagenStiftung;
the Department of Science and Technology of India;
the Istituto Nazionale di Fisica Nucleare of Italy;
National Research Foundation (NRF) of Korea Grant
Nos.~2016R1\-D1A1B\-01010135, 2016R1\-D1A1B\-02012900, 2018R1\-A2B\-3003643,
2018R1\-A6A1A\-06024970, 2018R1\-D1A1B\-07047294, 2019K1\-A3A7A\-09033840,
2019R1\-I1A3A\-01058933;
Radiation Science Research Institute, Foreign Large-size Research Facility Application Supporting project, the Global Science Experimental Data Hub Center of the Korea Institute of Science and Technology Information and KREONET/GLORIAD;
the Polish Ministry of Science and Higher Education and
the National Science Center;
the Ministry of Science and Higher Education of the Russian Federation, Agreement 14.W03.31.0026; 
University of Tabuk research grants
S-1440-0321, S-0256-1438, and S-0280-1439 (Saudi Arabia);
the Slovenian Research Agency;
Ikerbasque, Basque Foundation for Science, Spain;
the Swiss National Science Foundation;
the Ministry of Education and the Ministry of Science and Technology of Taiwan;
and the United States Department of Energy and the National Science Foundation.


%


\begin{thebibliography}{99}
\bibitem{theoC1}
M.~J.~Savage and R.~P.~Springer, Phys. Rev. D {\bf 42}, 1527 (1990).

\bibitem{theoC2}
M.~J.~Savage, Phys. Lett. B {\bf 257}, 414 (1991).

\bibitem{theoD1}
H.~Y.~Cheng, X.~W.~Kang, and F.~R.~Xu, Phys. Rev. D {\bf 97}, 074028 (2018).

\bibitem{theoD2}
J.~Zou, F.~Xu, G.~Meng, and H.~Y.~Cheng, Phys. Rev. D {\bf 101}, 014011 (2020).

\bibitem{theoD3}
W.~Wang, F.~S.~Yu, and Z.~X.~Zhao, Eur. Phys. J. C {\bf 77}, 781 (2017).

\bibitem{theoA1}
B.~Knapp {\it et al.}, Phys. Rev. Lett. {\bf 37}, 882 (1976).

\bibitem{pdg}
P.~A.~Zyla {\it et al.} (Particle Data Group), Prog. Theor. Exp. Phys. {\bf 2020}, 083C01 (2020).

\bibitem{theoB1}
P.~\.Zenczykowski, Phys. Rev. D {\bf 50}, 402 (1994).

\bibitem{theoB2}
K.~K.~Sharma and R.~C.~Verma, Phys. Rev. D {\bf 55}, 7067 (1997).

\bibitem{theoB3}
M.~A.~Ivanov, J.~G.~K$\rm \ddot{o}$rner, V.~E.~Lyubovitskij, and A.~G.~Rusetsky, Phys. Rev. D {\bf 57}, 5632 (1998).

\bibitem{theoB4}
Y.~Kohara, Nuovo Cim. A {\bf 111}, 67 (1998).

\bibitem{lhcbpomega}
R.~Aaij {\it et al.} (LHCb Collaboration), Phys. Rev. D {\bf 97}, 091101(R) (2018).

\bibitem{theoE1}
Y.~K.~Hsiao, Y.~Yao, and H.~J.~Zhao, Phys. Lett. B {\bf 792}, 35 (2019).

\bibitem{theoE2}
C.~Q.~Geng, C.~-W.~Liu, and T.~-H.~Tsai, Phys. Rev. D {\bf 101}, 053002 (2020).

\bibitem{theoF1}
P.~Singer and D.~-X.~Zhang, Phys. Rev. D {\bf 54}, 1225 (1996).

\bibitem{KEKB}
S.~Kurokawa and E.~Kikutani, Nucl. Instrum. Methods
Phys. Res., Sect. A {\bf 499}, 1 (2003), and other papers included
in this volume; T.~Abe {\it et al.}, Prog. Theor. Exp. Phys. {\bf 2013},
03A001 (2013), and references therein.

\bibitem{Belle}
A.~Abashian {\it et al}. (Belle Collaboration), Nucl. Instrum.
Methods Phys. Res., Sect. A  {\bf 479}, 117 (2002); also, see
detector section in J.~Brodzicka {\it et al.}, Prog. Theor. Exp. Phys. {\bf 2012}, 04D001 (2012).

\bibitem{pythia}
T.~Sj\"{o}strand, S.~Mrenna, and P.~Skands, Comput. Phys. Commun. {\bf 178}, 852 (2008).

\bibitem{evtgen}
D.~J.~Lange, Nucl. Instrum. Methods Phys. Res., Sect. A {\bf 462}, 152 (2001).

\bibitem{geant3}
R.~Brun {\it et al.}, CERN Report No. DD/EE/84-1, 1984.

\bibitem{topoana}
X.~Y.~Zhou, S.~X.~Du, G.~Li, and C.~P.~Shen, Comput. Phys. Commun. {\bf 258}, 107540 (2021).





\bibitem{pidcode}
E.~Nakano, Nucl. Instrum. Methods Phys. Res., Sect. A {\bf 494}, 402 (2002).

\bibitem{eidcode}
K.~Hanagaki, H.~Kakuno, H.~Ikeda, T.~Iijima, and T.~Tsukamoto, Nucl. Instrum. Methods Phys. Res., Sect. A {\bf 485}, 490 (2002).

\bibitem{speedoflight} We used units in which the speed of light is $c=1$.

\bibitem{ana-peta-ppi0}
S.~X.~Li {\it et al.} (Belle Collaboration), Phys. Rev. D {\bf 103}, 072004 (2021).

\bibitem{dalitz}
R.~H.~Dalitz, Phil. Mag. {\bf 44}, 1068 (1953).



\end{thebibliography}
\end{document}